# Quasilinear Evolution of Kinetic Alfven Wave Turbulence and Perpendicular Ion Heating in the Solar Wind

L. Rudakov[1], C. Crabtree, G. Ganguli and M. Mithaiwala

Plasma Physics Division, Naval Research Laboratory, Washington, DC 20375-5346
[1]Icarus Research Inc., P.O. Box 30780, Bethesda, MD 20824-0780 and
University of Maryand, Departments of Physics and Astronomy, College Park, Maryland 20742

## Abstract

The measured spectrum of kinetic Alfven wave fluctuations in the turbulent solar wind plasma is used to calculate the electron and ion distribution functions resulting from quasi-linear diffusion. The modified ion distribution function is found to be unstable to long wavelength electromagnetic ion cyclotron waves. These waves pitch angle scatter the parallel ion velocity into perpendicular velocity which effectively increases the perpendicular ion temperature.

## 1. Introduction

Recently the workshop on opportunities in plasma astrophysics (January 18-21, 2010, Princeton, New Jersey), endorsed by the APS topical group in plasma astrophysics, in reference to the kinetic Alfven waves (KAW) of scales larger than the proton gyro-radius $\rho_i$, concluded that: "When $k_\perp \rho_i > 1$, the ions decouple from the waves, and the damping is dominated by the electrons. As a result, the KAW do not undergo significant proton cyclotron damping in linear wave theory, but they do damp via Landau and transit-time damping. If KAW turbulence dissipates via Landau and transit-time damping, then the resulting turbulent heating should increase only the parallel component of the particle kinetic energy, thereby increasing the parallel temperature. On the other hand, in a number of systems such as the solar corona and solar wind, ions are observed to undergo perpendicular heating despite the fact that most of the fluctuation energy is believed to be in the form of low-frequency kinetic Alfvén wave fluctuations. *Determining the causes of such perpendicular ion heating is one of the critical unsolved problems in the study of space and astrophysical turbulence.*" In response to this assessment we suggest a possible physical process that could lead to a self-supportive mechanism of perpendicular ion heating.

Dispersive Alfvén waves have been considered as a possible candidate responsible for the solar wind turbulence [1, 2]. In this scenario the turbulent cascade of Alfven waves transfer energy from the inertial range consisting of scales larger than the proton gyro-radius $k_\perp \rho_i < 1$ to scales smaller than the gyro-radius $k_\perp \rho_i > 1$. In the inertial range the observed turbulent spectrum closely follows the Kolmogorov scaling of $1/k^{5/3}$ [3, 4]. At smaller scales the wave vector spectrum of the turbulence is highly anisotropic with energy concentrated in wave vectors nearly perpendicular to the mean magnetic field $B_0$ such that $k_\perp \gg k_\parallel$. It was recognized that electron Landau damping affects the scaling of the turbulent spectra in the range $k_\perp \rho_i > 1$ [5]. The damping of KAW computed by assuming a Maxwellian distribution is too strong to create a steepened power spectrum as observed but rather leads to sharp cutoffs [6-8].

We have recently suggested that the Landau damping of KAW leads to the quasi-linear evolution of the parallel electron velocity distribution function which diminishes the Landau damping and enables unique non-linear plasma dynamics [9]. In the solar wind the Landau damping time of KAW for a Maxwellian distribution function is a few tenths of a second while the time of flight of the solar wind to the earth is $10^5$ s. Additionally, the Coulomb electron collision time and mean free path in the "fast" solar wind plasma with density ~10/cc and temperature ~10 eV is also about $10^5$ s and a few AU respectively. The proton collision time is about $10^7$ s [3]. Hence an initial Maxwellian distribution can evolve by quasi-linear diffusion and remain non-Maxwellian for its lifetime in the solar wind. The diffusion time depends on the established turbulence level. The measured spectrum is used to determine the rate of diffusion and the analytical form of the distribution function. It is found that the diffusion rate is consistent with the time of flight of solar wind plasma to the Earth.

**2. KAW in solar wind plasma - Linear theory**

In a high beta plasma with Maxwellian distribution for ions and electrons the dispersion relation for KAW with $\vec{E} = \vec{E}_k \exp(-i\omega t + ik_x x + ik_z z)$, $\omega \ll \Omega_i$ and $\omega/k_\parallel v_{te} < 1$ is [1,2]

$$\omega^2 = \frac{k_\parallel^2 \left((T_i + T_e)/m\right)}{1 + (\beta_i + \beta_e)/2} \frac{k_\perp^2 c^2}{\omega_{pe}^2}. \tag{1}$$

Here $k_\parallel \equiv k_z$, $k_\perp \equiv k_x$ are the wave vector components along and normal to the magnetic field, $\beta_s \equiv 8\pi n_0 T_s / B_0^2$, and the temperature $T_s \equiv m_s v_{ts}^2 / 2$ for thermal speed $v_{ts}$. The gyro-radius will be denoted $\rho_s \equiv v_{ts}/\Omega_s$ with cyclotron frequency $\Omega_s \equiv eB_0/m_s c$ for each species (where $s = e$ or $i$ for electrons or ions respectively), and the mass by $m$ for electrons and $M$ for ions. Recently using space-correlated measurements made by four Cluster satellites in the solar wind it was shown that for $k_\perp \rho_i \approx 2$ the representative

KAW in the turbulent spectra have $\omega \sim 0.1\Omega_i$ with $k_\parallel / k_\perp \sim 1/10$ [10]. Morphologically KAWs resemble elongated filaments along $B_0$.

The electric field for KAWs with electron and ion distribution functions that are arbitrary in $v_z$ and Maxwellian in $v_\perp$ can be obtained from Ampere's law $\vec{\nabla} \times \vec{B} = 4\pi\vec{j}/c$ along with $\vec{B} = \vec{\nabla} \times \vec{A}$, $\vec{\nabla} \cdot \vec{A} = 0$ and $\vec{E} = i\omega\vec{A}/c - i\vec{k}\varphi$. Eliminating the vector-potential $\vec{A}$ from the time derivative of Ampere's law we obtain the equation which expresses the electric field vector $\vec{E}$ in terms of the electrostatic potential $\varphi$ [9],

$$\bar{k}^2(\vec{E}+\vec{\nabla}\varphi) = \int \frac{\omega v_z dv_z}{\omega - k_z v_z} \frac{\partial f_{0e}}{n_0 \partial v_z}(\vec{E}_z + \frac{\mu_e}{e}\vec{\nabla}_z B) + \frac{i\omega}{\Omega_e}(\vec{E}\times\vec{b}) - \frac{\omega}{\Omega_e}(\vec{k}\times\vec{b})\frac{T_{\perp e}\delta n}{n_0}$$
$$+ \frac{2m\omega^2 \vec{E}_x}{Mk_\perp^2 v_{ti}^2} + \frac{m}{M\pi^{1/2}(k_\perp \rho_i)} \frac{i\pi\omega^2}{k_z^2} \frac{\partial f_{0i}(v_z = \omega/k_z)}{n_0 \partial v_z}\vec{E}_z \quad (2)$$

Here $\bar{k}^2 = k^2 c^2 / \omega_{pe}^2$, $k_\perp^2 \rho_i^2 \gg 1$, $T_{\perp s}$, $T_{\parallel s}$ and $\mu_e = mv_\perp^2 / B$ are the perpendicular and parallel particle temperatures and electron magnetic moment. It is assumed that $T_{\perp e} = T_{\parallel e}$. The third term on the right hand side (RHS) of Eq. (2) is due to the electron diamagnetic current $\vec{j}_e = -c(T_{\perp e}\vec{\nabla}\delta n \times \vec{b})/B_0$. In the solar wind plasma $\beta_i \sim \beta_e \sim O(1)$. However, for simplicity of calculations we assume that $m/M \ll (\beta_i, \beta_e) < 1$. A significant advantage of this approximation is that the KAW is distinct from magnetosonic waves. For an isothermal plasma, $T_{\perp e} = T_{\parallel e}$, the estimates obtained under this assumption can be extended up to $\beta \sim 1$.

For a low beta plasma and highly oblique ($k_\parallel^2 / k_\perp^2 \ll 1$) KAW the parallel and perpendicular components of the electric field (2), magnetic field $B_y$ and energy density $W_k$ yield [9]

$$E_z = -ik_z \varphi \left( 1 + \int \frac{\omega^2 dv_z}{\overline{k}^2 k_z^2 (v_z - \omega/k_z)} \frac{\partial f_{0e}}{n_0 \partial v_z} \right)^{-1}, \quad (3)$$

$$\vec{E}_\perp = -i\vec{k}_\perp \varphi \left( 1 + O(k_\parallel^2 / k_\perp^2) \right), \quad (4)$$

$$\frac{|B_y|^2}{B_0^2} = \frac{\beta_i + \beta_e}{2} \frac{|e\varphi|^2}{T_i^2} = \frac{W_k}{B_0^2 / 4\pi}. \quad (5)$$

In the short perpendicular wavelength electric field of KAW (Eq. (4)) ions are effectively unmagnetized, their density perturbation can be described by the Boltzmann distribution $\delta n / n = -e\varphi / T_{\perp i}$, and there is no cyclotron damping. However ions oscillate in the parallel electric field $E_z$ and can satisfy the Landau resonance condition and damp the KAW as indicated in the last term in the RHS of Eq. (2).

**3. Quasi-linear evolution of particle distribution functions in solar wind**

The quasi-linear equations [11] for magnetized plasmas in the limit $\omega_k < \Omega_s$ relevant for KAW

$$\frac{\partial f_{0s}}{\partial t} = \frac{\partial}{\partial v_z} D_s(v_z, v_\perp) \frac{\partial f_{0s}}{\partial v_z}, \quad (6)$$

$$D_s(v_z, v_\perp) = \frac{\pi e^2}{m_s^2} \int |E_{kz}|^2 J_0^2(\lambda_{ks}) \delta(\omega_k - k_z v_z) d\vec{k}, \quad (7)$$

determines the slow evolution of the background distribution function $f_{0s}(t, v_z)$ in a turbulent plasma as a result of multiple radiation and absorption of plasmons through the Landau resonance. $\lambda_{ks} = k_\perp v_\perp / \Omega_s$, $J_0(\lambda_{ks})$ is the Bessel function, and the electron and ion distribution functions over $v_\perp^2$ initially are Maxwellians. For electrons $\lambda_{ke} \ll 1$, $J_0^2(\lambda_{ke}) \cong 1$, and for ions, $\lambda_{ki} \gg 1$, $J_0^2(\lambda_{ki}) \cong 2\cos^2\lambda_{ki}/\pi\lambda_{ki}$. To estimate $D_s$ we use the measurements from Ref. [3] and [9], i.e., $B_0 \sim 10 nT$, $c/\omega_{pe} \sim 1.5\, km$ and, as in Fig. 1 (reproduced from ref. [3]), $P(k_\perp)/P_0 \sim 10^{-4}/\bar{k}_\perp^\nu$, $\nu = 2.8$ (we use 3 instead 2.8 for simplicity). As mentioned above, the KAW frequency for $k_\perp \rho_i = 2$ was measured to be $0.1\Omega_i \sim 0.1/s$ [10]. For larger $k_\perp \rho_i$ the frequency could be different, however we estimate the average frequency to be $\langle \omega_k \rangle \sim 0.1/s$.

Asymptotically Quasi-linear diffusion Eq. (6) establishes and maintains a step-like distribution in the velocity range $V_A < |v_z| < v_{ms}$ with maximum velocity $v_{ms}$ and density $\alpha_{ms} n_0$ [9]. The width of the plateau in the parallel velocity grows in time at a rate that depends on the turbulence amplitude. Figure 2a shows the model electron distribution function. In the interval $|v_z| < V_A = B_0/(4\pi n_0 M)^{1/2}$, the cold plasma has a characteristic thermal velocity $v_{cs}$ and density $\alpha_{cs} n_0$. In the interval $|v_z| > v_{me}$, the electron Maxwellian has a characteristic thermal velocity $v_{te}$ and density $\alpha_{te} n_0$, where $n_0$ is the total density.

Since the electron distribution function is no longer entirely Maxwellian the dispersion relation for KAW and $E_{kz}$ and $B_{k\perp}$ are modified. In the limit $1/\rho_i^2 \ll k^2 \ll \omega_{pe}^2/c^2$ we obtain from Eqs. (2-5)

$$\omega^2 = k_z^2 \bar{k}_\perp^2 v_0^2, \tag{8}$$

$$E_{kz} \sim ik_z \varphi (mv_e^2/T_{\perp i}), \tag{9}$$

$$\frac{|\delta B_{k\perp}|^2}{B_0^2} \sim \frac{e^2|\varphi_k|^2}{T_{\perp i}^2}\frac{mv_0^2}{MV_A^2}, \tag{10}$$

where $v_0^2 = (v_e^2 + T_{\perp i}/m)$, $v_e^2 \sim \frac{(\alpha_t + \alpha_{me})v_{me}^2 v_{te}^2}{\alpha_t v_{me}^2 + \alpha_{me}v_{te}^2}$ is obtained by interpolation between the Maxwellian and step distributions, and $\bar{k} = kc/\omega_{pe}$.

First we analyze the evolution of electron distribution function. The width of the plateau distribution, determined by $v_{me}$, increases with time. To ensure that the plateau can be created within the time of travel of solar wind plasma to the Earth we must solve the diffusion equation. Using Eqs. (9), (10) the diffusion coefficient (7) can be written as

$$D_e = \frac{\pi MV_A^2}{m}\int\frac{\langle\omega_k\rangle v_e^4}{\bar{k}_\perp v_0^4}\frac{P(k_\perp)/P_0}{B_0^2(c/\omega_{pe})}\delta(\bar{k}_\perp - |v_z|/v_0)d\bar{k}_\perp \sim \frac{2\times 10^{-7}}{(v_z/v_e)^4}\frac{MV_A^2}{m}. \tag{11}$$

It is assumed that the turbulence is isotropic in $\bar{k}_\perp$ and the electric field (9) is expressed through the measured magnetic fluctuation spectrum, $P(k_\perp)/P_0 \sim |\delta B_{k\perp}|^2/k_\perp$. Analysis of Eqs. (6), (11) gives the estimate that for $t \sim 10^5 s$ the size of the plateau is $v_{me}/v_{te} \sim (\beta_e 10^{-7} t)^{1/6} \sim 0.5$. A numerical solution of the diffusion equation (6) beginning with a Maxwell distribution function with $D_e \sim v_z^{-4}$ (11) is shown in Fig 2b. There is good agreement between the model distribution Fig2a [9] and the time asymptotic solution of the diffusion equation.

Evaluation of $f_{0i}(v_z, t)$ is not as straight forward because the minimum KAW phase velocity is $V_A \sim v_{ti}$, which is in the exponentially small tail of the distribution, and only a small fraction of the bulk ions $\alpha_{ci}$ are involved in quasi-linear heating. This is contrary to the electrons where the bulk of the distribution is involved in the quasi-linear heating. The model ion distribution function is shown in figure 3. For the interval $v_\parallel < V_A \sim v_{ti}$, KAW cannot meet the phase velocity resonance with ions, and a quasi-Maxwellian distribution remains. The ion densities over these velocity intervals are $\alpha_{ci} n_0$, $\alpha_{mi} n_0 = (1-\alpha_{ci}) n_0$.

Additionally, near the transition from regular shear Alfven waves with $\omega = k_z V_A$ for $k_\perp \rho_i < 1$ to KAW of Eq. (8) at $k_\perp \rho_i > 1$ there is a discontinuity in the derivative of the amplitude $P(k_\perp)/P_0 \sim |\delta B_{k\perp}|^2/k_\perp$ as seen in Fig. 1 [3, 9] implying additional physics, which is not critical for the electrons because the resonance is in the bulk of the electron population. Hence $D_i$ is not expressible as a simple function of $v_z^{-n}$ as in the electrons (11). However as we shall see the distribution of the ions that are in exponential tail expands in quasilinear diffusion along $v_z$. Ignoring the diffusion of ions across the boundary $k_\perp \rho_i = 1$ (which can be addressed numerically), it is possible to find asymptotic self-similar solution of the ion diffusion equation for $k_\perp \rho_i > 1$.

The ion diffusion coefficient $D_i(v_z) \propto J_0^2(\lambda_k)$ where $\lambda_k = k_\perp v_\perp/\Omega_i$. The resonance condition $(k_\perp c/\omega_{pe}) = |v_z|/v_0$ determines $k_\perp$. This makes $\lambda_k \sim v_\perp |v_z| \beta_e^{1/2} M/m v_0^2$ dependent on $v_z$ as well as magnetic field and plasma parameters. Consequently diffusion over $v_z$ in $(v_\perp, v_z)$ space vanishes where $J_0^2(\lambda_k) = 0$. This leads to the formation of multiple plateaus in $(v_\perp, v_z)$ space with sharp boundaries and under uniform conditions the ions will be confined within each plateau. But on the time scale of solar wind dynamics the resonant ions move along magnetic field lines over a distance $\sim 10^6$ km. Stochastic changes in the magnetic field and plasma density over this distance as in long wavelength part on Fig. 1 change $\lambda_k$ stochastically as well. This enables the ions to jump across the plateau boundaries in the velocity space to experience an average diffusion over its lifetime. The ion diffusion trajectories in $(v_\perp, v_z)$ space are shown schematically in Fig. 3a. Estimating the average over such fluctuations, $\langle J_0^2(\lambda_k) \rangle \sim 1/k_\perp \rho_i$, the ion diffusion coefficient is similar to the electron diffusion coefficient (11)

$$D_i(v_z) \sim \frac{\pi^{1/2} M V_A^2}{m} \int \frac{\langle \omega_k \rangle v_S^4}{\bar{k}_\perp v_0^4 (k_\perp \rho_i)} \frac{P(k_\perp)/P_0}{B_0^2 (c/\omega_{pe})} \delta(\bar{k}_\perp - |v_z|/v_0) d\bar{k}_\perp \sim \frac{4 \times 10^{-4} v_S^3}{|v_z/v_S|^5 v_\perp \beta_e^{3/2}}, \quad (12)$$

where $v_S^2 = m v_e^2 / M \sim 2 T_e / M$.

The asymptotic solution of Eq. (6) is obtained using the *ansatz* that $f_{0i}(v_z, t)$ is a self-similar function of $v_z/t^{1/7}$. This solution conserves the number of ions that participate in quasilinear diffusion. Substituting $f_{0i}(v_z/v_S (4 \times 10^{-4} \sigma t)^{-1/7})$ in Eq. (6) we get

$$f_{0i}(|v_z| > V_A) \sim \frac{\alpha_{mi} n_0}{2 v_{mi}} \exp\left(-\frac{|v_z|^7}{v_{mi}^7}\right) \quad (13)$$

where $v_{mi} \sim v_S (\tau_i \beta_e^{3/2} v_S / v_\perp)^{1/7}$ and $\tau_i = 4 \times 10^{-4} \sigma t(s)$. The constant $\sigma = 7^2$ is determined upon substituting (13) in (6). Dependence of $v_{mi}$ on $v_\perp$ as well as from $\beta_e$ is very weak and to simplify calculations we use $v_{mi} \sim v_S (\tau_i)^{1/7}$ since $(\beta_e^{3/2} v_S / v_\perp)^{1/7} \sim O(1)$. Then at the normalized time $\tau_i \sim 10^3 (t \sim 10^5 s)$, we estimate $v_{mi}^2 \sim 10 v_S^2$, the ion energy at the plateau front as $M v_{mi}^2 / 2 \sim 10 T_e$. In Fig. 3b the analytical solution (13) is compared with a numerical solution of the diffusion equation (6) where $D_i \sim v_z^{-5}$ (12).

The electron Coulomb collision time in the fast solar wind plasma is about $10^5$ s and the proton collision time is about $10^7$ s. The electron distribution function can be partially thermalized by Coulomb collisions during the solar wind time of flight toward earth $\sim 10^5$ s. But plateau remains in the interval $V_A < |v_z| < v_{me} < v_{te}$ with a smooth front because the rate of quasi-linear diffusion $\sim v_z^{-6}$ while Coulomb relaxation rate $\sim v^{-3}$. However the impact of collisions on the ion distribution function is negligible.

**4. Cyclotron instability of the ion step-like distribution function**

The sharp front in the ion distribution function developed due to the quasi-linear evolution can be the source of instabilities. We consider the instability of the left hand polarized electro-magnetic ion cyclotron (EMIC) waves in the solar wind plasma with the ion distribution function given in Eq. (13). As was shown by Sagdeev and Shafranov [12], and then in numerous papers that distribution function anisotropy, in particular the temperature anisotropy $T_\perp > T_\parallel$, can drive such an instability. In our case the ion distribution function has $\alpha_{ci} n_0$ "cold" particles with $v_z < V_A$ and a small amount $\alpha_{mi} n_0$ of "warm" particles with non-Maxwellian distribution function for $|v_z| > V_A$ given by Eq. (13). The dispersion relation for EMIC waves propagating along the magnetic field with $k_\perp \ll k_z$ is given by [11, 13],

$$\frac{k_z^2 c^2}{\omega_{pi}^2} = \frac{\alpha_c \omega}{2(\Omega_i - \omega)} + \pi \omega \int dv_z v_\perp^2 dv_\perp \left( \frac{\partial f_{0i}}{\partial v_\perp} - \frac{k_z v_z}{\omega} \frac{\partial f_{0i}}{\partial v_\perp} + \frac{k_z v_\perp}{\omega} \frac{\partial f_{0i}}{\partial v_z} \right) \frac{1}{\omega - \Omega_i - k_z v_z}. \quad (14)$$

After integration over $v_\perp$, using a Maxwellian distribution function over $v_\perp$ the dispersion relation (14) becomes

$$\frac{k_z^2 c^2}{\omega_{pi}^2} = \frac{\alpha_{ci} \omega}{2(\Omega_i - \omega)} - \Omega_i \int \frac{dv_z}{\omega - \Omega_i - k_z v_z} \left( f_{0i} + \frac{k_z}{\Omega_i} \frac{v_{\perp i}^2}{2} \frac{\partial f_{0i}}{\partial v} \right). \quad (15)$$

After integrating over $v_z$ with $f_{0i}$ from (13) we separate the real and small imaginary parts of Eq. (15), which appears due to cyclotron resonance $\Omega_i - \omega = -k_z v_z$,

$$\frac{k_z^2 c^2}{\omega_{pi}^2} = \frac{\alpha_{ci} \omega}{2(\Omega_i - \omega)} \qquad (16)$$

$$\gamma = -\frac{\pi \alpha_{mi}(\Omega_i - \omega)^2}{2|k_z|v_{mi}} \left(1 - \frac{(\Omega_i - \omega)}{\Omega_i} \frac{7 v_{\perp i}^2 |v_z|^7}{2 v_z^2 v_{mi}^7}\right) \exp\left(-\frac{|v_z|^7}{v_{mi}^7}\right), \qquad (17)$$

where $|k_z v_z| = \Omega_i - \omega$, $v_{mi} \sim v_S \left(10^{-2} t(s)\right)^{1/7}$, and $v_S^2 = 2T_e/M \sim V_A^2$.

For the solar wind plasma, where $\Omega_i \sim 1/s$, $c/\omega_{pi} \sim 10^7 \, cm$, and $v_S \sim V_A \sim 5 \times 10^6 \, cm/s$, the resonance condition $\Omega_i - \omega = -k_z v_z$ can be met for $\beta_i (v_z/v_S)^2 \sim (\Omega_i/\omega) \gg 1$. The EMIC wave instability exists for $\omega < \Omega_i$ in a broad range of the plasma parameters and anisotropy ratio $\theta = v_{\perp i}^2 / v_{mi}^2$ even for small $\theta$. For example for $|v_z|^7 / v_{mi}^7 = 3$ there is instability for $\theta > 1/10$ and we estimate the growth rate as

$$\gamma \sim 0.01 \alpha_{mi} \Omega_i \sim 10^{-3}/s. \qquad (18)$$

The EMIC waves can scatter ions over pitch-angle. The pitch-angle scattering rate by EMIC waves is estimated as $v_{\vartheta\vartheta} \sim \Omega_i |\delta B_\perp|^2 / B_0^2$. The turbulent magnetic field energy in EMIC waves $|\delta B_\perp|^2 / B_0^2 \sim 10^{-5}$ is enough to scatter the ions in $10^5$ second, which is the approximate time of flight of solar wind plasma to the Earth. On Fig. 1 the arrow indicates the possible position of EMIC waves where measured magnetic field energy is $|\delta B_\perp|^2 / B_0^2 \sim 10^{-2}$. Thus with less than 1% of turbulent magnetic field energy in EMIC waves a quasi-equilibrium with $T_{\perp i} \sim T_{\|i}$ can be supported.

Contrary to KAW, most of the unstable EMIC waves have small $k_\perp/k_\parallel$ and consequently experience a small Doppler shift as compared to KAW. Satellites in fast solar wind with $V_{SW} \sim 600 km/s$ measure the Doppler shifted frequency. For KAW, $\omega_k \ll k_\perp V_{SW}$ and satellites effectively measure the wave vector $k_\perp$ [3], while for EMIC satellites would measure the frequency. This may allow us to identify the EMIC contribution in the net wave spectrum that is observed.

We have shown that the evolution of ion distribution function in turbulent solar wind plasma can lead to the generation of electromagnetic ion cyclotron waves. These waves pitch-angle scatter the super-thermal ions. Isotropization of the plateau-like ion distribution by the pitch angle scattering involves scattering the parallel ion velocity into perpendicular velocity which is effectively ion heating. To determine the efficiency of this process self-consistent numerical simulations are necessary. The analytical framework provided here can be the basis to develop an appropriate numerical simulation model.

**Acknowledgments**

This work is supported by the Office of Naval Research.

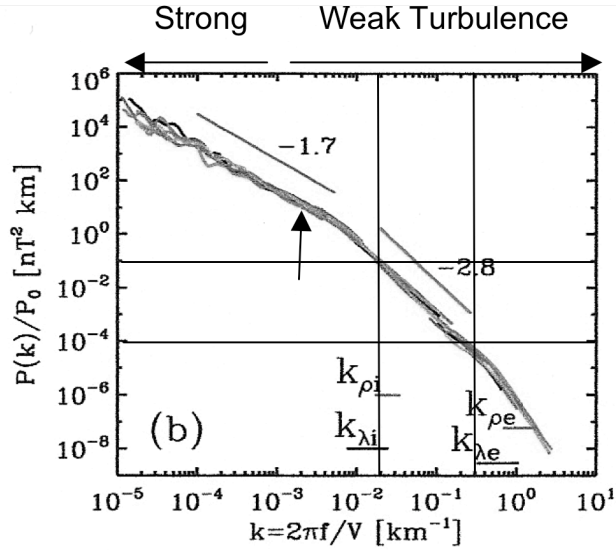

Figure 1. The solar wind magnetic field spectrum from Cluster data [3]. Here $k_{\rho i,e} \sim 1/\rho_{i,e}$, $k_{\lambda i,e} \sim \omega_{pi,e}/c$, $P(k_\perp)/P_0 \sim 10^{-4}(kc/\omega_{pe})^{-\nu}$. With less than 1% of turbulent magnetic field energy in EMIC waves a quasi-equilibrium with $T_{\perp i} \sim T_{\parallel i}$ can be supported. The arrow indicates the possible position of EMIC waves.

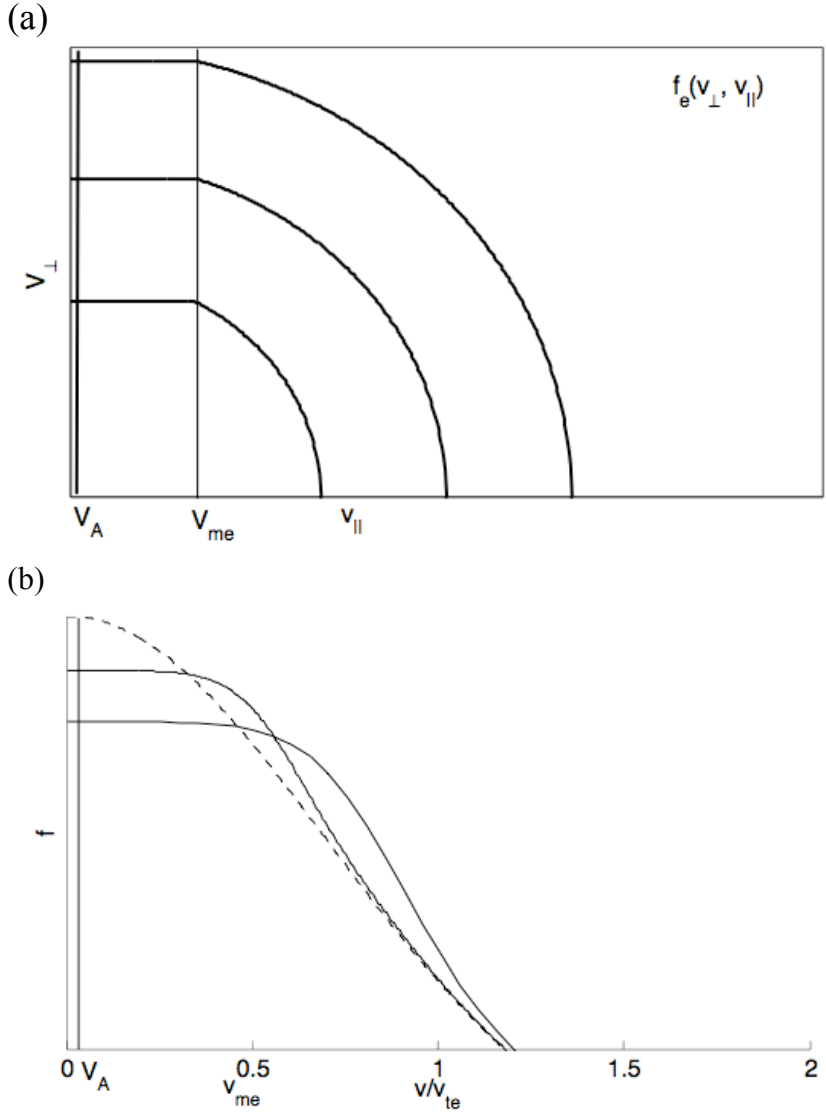

**Figure 2. a)** Level curves of the model distribution electron distribution function. Quasi-linear diffusion establishes a plateau in the velocity range $V_A < v_\| < v_{me}(t)$. Outside this interval the distribution remains Maxwellian. **b)** A numerical solution of the diffusion equation (6), beginning with a Maxwell distribution function (---) with $D_e \sim v_z^{-4}$ (11), and evolving to times $t_1 = 10^4 s$ and $t_2 = 10^5 s$.

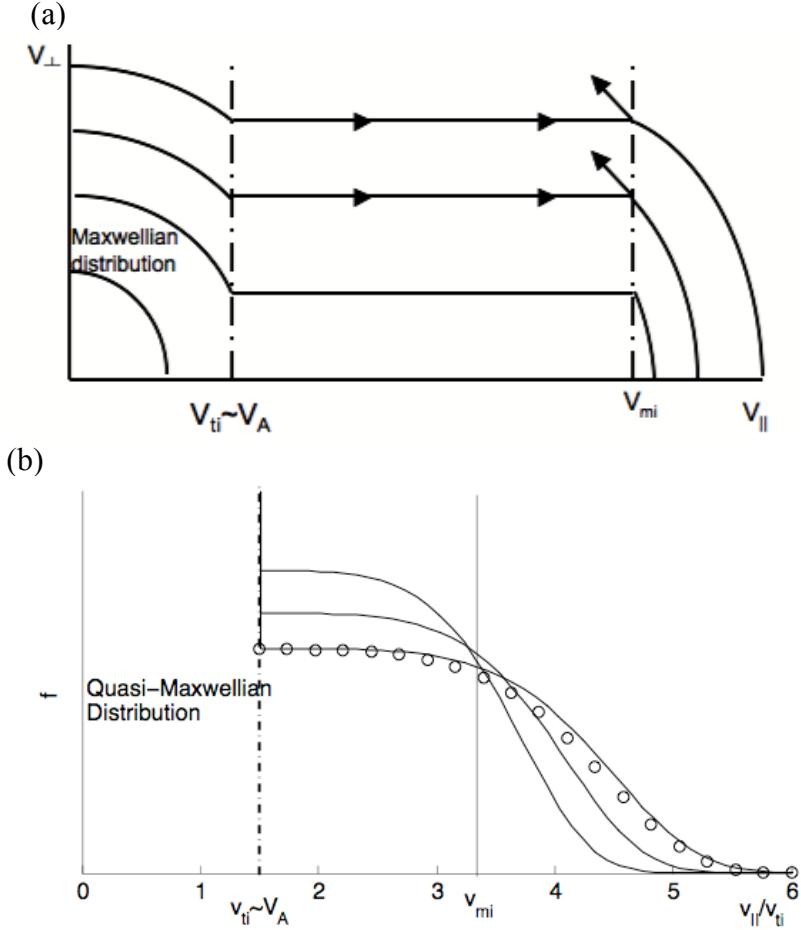

**Figure 3. a)** Level curves of the model distribution ion distribution function. The front of the ion distribution with large $\partial \ln f_{0i}/\partial v$ is unstable to EMIC waves that pitch angle scatter ions from parallel to perpendicular velocities effectively heating the plasma. The right-pointing arrows indicate the direction of quasi-linear and the left-pointing arrows indicate pitch-angle diffusion. **b)** Numerical solution of the diffusion equation (6) beginning from an initial Maxwell distribution function (not shown) with $D_i \sim v_z^{-5}$ (12). In the interval $v_\parallel < V_A \sim v_{ti}$, KAW cannot meet the phase velocity resonance with ions, and a quasi-Maxwellian distribution remains. The analytical solution (marked by open circles 'o') of Eq. 13 is compared with the numerical solution and they are found to be in good agreement.